\documentclass[conference]{IEEEtran}

\IEEEoverridecommandlockouts
\usepackage{cite}
\usepackage{amsmath,amssymb,amsfonts}
\usepackage{algorithm}
\usepackage{algorithmicx}
\usepackage[noend]{algpseudocode}
\newcommand{\codecomment}[1]{\textcolor{gray}{\# #1}}
\algdef{SE}[SUBALG]{Indent}{EndIndent}{}{\algorithmicend\ }
\algtext*{Indent}
\algtext*{EndIndent}

\usepackage{placeins}
\usepackage{graphicx}
\usepackage{textcomp}
\usepackage[table,xcdraw]{xcolor}
\usepackage{hhline}
\usepackage{enumitem}
\usepackage{booktabs}
\usepackage[caption=false]{subfig}

\usepackage[utf8]{inputenc}
\graphicspath{{./figs/}}
\usepackage{graphicx, epstopdf}
\usepackage{float}

\usepackage{ifthen}
\usepackage{array}
\usepackage{makecell}
\usepackage{pdfpages}
\usepackage{mathtools}
\usepackage{tikz}
\usepackage{multirow}
\usepackage{booktabs}
\usepackage{multirow}
\usepackage{tabularx}
\usepackage{svg}

\usepackage{hyperref}
\usepackage{cleveref}

\newcolumntype{?}{!{\vrule width 1pt}}

\newlength\myindent
\makeatletter
\newcommand{\multiline}[1]{%
  \begin{tabularx}{\dimexpr\linewidth-\ALG@thistlm}[t]{@{}X@{}}
    #1
  \end{tabularx}
}
\makeatother

\DeclareMathOperator{\nnz}{nnz}

\title{Fast multiplication of random dense matrices with sparse matrices}

\author{
\IEEEauthorblockN{Tianyu Liang\IEEEauthorrefmark{1}\IEEEauthorrefmark{2}, Riley Murray\IEEEauthorrefmark{1}\IEEEauthorrefmark{2}\IEEEauthorrefmark{3}\IEEEauthorrefmark{4}, Ayd{\i}n Bulu\c{c}\IEEEauthorrefmark{2}\IEEEauthorrefmark{1}, James Demmel\IEEEauthorrefmark{1}\IEEEauthorrefmark{2}}
    \IEEEauthorblockA{\IEEEauthorrefmark{1}Electrical Engineering and Computer Science Department, UC Berkeley}
    \IEEEauthorblockA{\IEEEauthorrefmark{2}Computational Research Division, Lawrence Berkeley National Lab
    }
    \IEEEauthorblockA{\IEEEauthorrefmark{3}International Computer Science Institute
    }
    \IEEEauthorblockA{\IEEEauthorrefmark{4}Sandia National Laboratories
    }
}

\begin{document}

\maketitle
\thispagestyle{plain}
\pagestyle{plain}

\begin{abstract}
This work focuses on accelerating the multiplication of a dense random matrix with a (fixed) sparse matrix, which is frequently used in sketching algorithms. We develop a novel scheme that takes advantage of blocking and recomputation (on-the-fly random number generation) to accelerate 
this operation.
The techniques we propose decrease memory movement, thereby increasing the algorithm's parallel scalability in shared memory architectures. On the Intel Frontera architecture, our algorithm can achieve 2x speedups over libraries such as Eigen and Intel MKL on some examples. In addition, with 32 threads, we can obtain a parallel efficiency of up to 45\%.

We also present a theoretical analysis for the memory movement lower bound of our algorithm, showing that under mild assumptions, it's possible to beat the data movement lower bound of general matrix-matrix multiply (GEMM) by a factor of $\sqrt{M}$, where $M$ is the cache size. Finally, we incorporate our sketching method into a 
randomized algorithm for overdetermined least squares with sparse data matrices.
Our results are competitive with SuiteSparse for highly overdetermined problems; in some cases, we obtain a speedup of 10x over SuiteSparse.
\end{abstract}

\section{Introduction}
\subsection{Background}
In this work, we explore methods for accelerating dense-sparse matrix multiply (SpMM), where the dense matrix is random.
In particular, given a tall sparse matrix $A \in \mathbb{R}^{m \times n}$ (tall meaning $m \gg n$) and a parameter $d$ only slightly larger than $n$ (say, $n < d \leq 4n$), we want a fast kernel for sampling a random matrix $S \in \mathbb{R}^{d \times m}$ from a prescribed probability distribution and computing the matrix matrix product:
\begin{equation}
    \hat{A} = SA.
    \label{goal}
\end{equation}
This computational primitive is a particular type of \textit{sketching},
which has applications in randomized algorithms for linear regression, low-rank approximation, matrix decomposition, eigenvalue computation, and many more tasks \cite{murray2023randomized}.
Our work on this computational kernel is motivated by a desire to help realize the full potential of such randomized algorithms.
As a matter of scope, we are concerned with the case when the entries of $S$ are independent and identically distributed; the underlying distribution can be mean-zero Gaussian, uniform over $(-1, 1)$, or simply uniform over $\pm 1$.
We focus more on the latter distributions since they are much
cheaper to sample from, and we take CSC as our default sparse matrix format.

SpMM operations have been studied extensively by many. Some researchers \cite{nuclear, CSB} designed blocked sparse data structures to improve performance. Other approaches sought to add an analysis stage before computation, allowing the algorithm to choose the most appropriate method. Intel's ``inspector-executor" model in the MKL sparse library \cite{intel} and matrix signature estimation \cite{matrix:signature} are examples of such. Researchers also developed advanced tools for compile or run time analysis\cite{automatic_parallelization, optimizing_irregular_Shared, Loop_and_data}. In addition, vector instructions (e.g., AVX) prove to be effective at accelerating computations \cite{liu, bramas2018computing, dynamical}. 

\subsection{Our Contribution}
In this paper, we develop a novel scheme that takes advantage of blocking and on-the-fly random number generation to accelerate SpMM involving random matrices.
This technique improves parallel scalability by decreasing memory movement. We explore the design space that arises when considering this approach, we report results of numerical experiments with Julia implementations of several methods in this category, and we present theoretical analysis to understand the potential of this technique.
The analysis shows that our methodology can beat the data movement lower bound of general matrix-matrix multiply (GEMM) by a factor of $\sqrt{M}$ ($M$ is the cache size) in a suitable data model.

Finally, we use our sketching method in a full ``pipeline'' of a randomized algorithm for solving least squares problems:
\begin{equation}
    \min_x \|Ax - b\|_2. \label{eq:least_square}
\end{equation}
For extremely tall and sparse $A$, our randomized solver is competitive with SuiteSparseQR \cite{suitesparse} (accessed via Julia's backslash solver) in terms of runtime and accuracy, and attains dramatic savings over SuiteSparse in terms of workspace requirements.
This last point warrants emphasis: our randomized least squares solver needs to compute and factor a \textit{dense} matrix $\hat{A} = SA$, and yet in many cases has lower memory requirements than a direct sparse solver. Our implementation can be found here.\footnote{\url{https://github.com/Tianyu-Liang/Solving-with-sketch}}

\section{Multiplication Algorithm}
In this section, we divide the discussion into outer blocking schemes and inner compute kernels. Outer loops correspond to the methodology for blocking/partitioning the matrices, while compute kernel refers to the procedure that we use to actually perform the computations on the partitioned blocks.

\subsection{Outer Blocking Scheme}
We can use a 3-tuple to define an outer blocking; the three numbers represent the number of block rows of $\hat{A}$ and $S$, the number of block columns of $S$ and block rows of $A$, and the number of block columns of $\hat{A}$ and $A$, respectively.
For example, here is a depiction of $(3, 1, 2)$-blocking:
\begin{align}
    SA = \begin{pmatrix}
       S_1 \\ S_2 \\ S_3
    \end{pmatrix}\begin{pmatrix}
       A_1 & A_2 
    \end{pmatrix} &= \begin{pmatrix}
       S_1A_1 & S_1A_2 \\ S_2A_1 & S_2A_2  \\ S_3A_1 & S_3A_2
    \end{pmatrix}.  \label{simpleblock}
\end{align}

To find the optimal blocking scheme we must consider our approaches for forming $S$.
The naive approach is to generate $S$ beforehand and then call library routines such as Intel MKL to perform SpMM. 
This approach is not practical for large inputs because $S$ may not even fit into RAM.
Indeed, in our context of sketching sparse matrices, it has recently been proposed to generate the columns of $S$ only one at a time \cite{https://doi.org/10.48550/arxiv.2203.02798}.
Such an approach (corresponding to $(1, m, 1)$-blocking, in our parlance) is memory efficient, but it is not cache friendly since the outer product will update the entirety of $\hat A$.

In summary, we want to find a blocking scheme that can both reduce memory usage and exploit cache locality. The key to achieving this goal is to realize that generating random numbers can be faster than moving entries from main memory into cache.
That is, we can use on-demand random number generation to convert a portion of memory movement cost into computation cost.

\begin{algorithm}[b]
    \caption{Loop blocking corresponding to \cref{simpleblock}} \label{alg:simple}
    \begin{algorithmic}[1]
        \Require{\multiline{
                $A$\,: sparse $m \times n$ matrix,
            \\ $  d$\,: sketch size, 
            \\ $b_n$\,: block size of the side with length $n$,
            \\ $b_d$\,: blocking size of the side with length $d$
            }
        }
        \State $\hat{A} = 0_{\mathbb{R}^{d \times n}}$
        \For{$j = 1 \texttt{:} b_n \texttt{:} n$} 
            \State n\_stop = $\min(n, j + b_n - 1)$
            \For{$i = 1 \texttt{:} b_d \texttt{:} d$}
                \State d\_stop = $\min(d, i + b_d - 1)$                    
                \State{
                    \multiline{
                        $\texttt{compute\_kernel(}$ \\
                        $\quad \hat{A}[i \texttt{:} \text{d\_stop}, j \texttt{:} \text{n\_stop}],~ A[\texttt{:},\,j \texttt{:}\text{n\_stop}], ~i$ \\
                        $\texttt{)}$ \codecomment{transform $\hat{A}$ in-place}
                    } 
                }\label{subblock_index}      
            \EndFor
        \EndFor
        \State \textbf{return} $\hat{A}$
    \end{algorithmic}
\end{algorithm}

 \Cref{alg:simple} generalizes the outer loop blocking pattern shown in \Cref{simpleblock}. It uses $(\lceil \frac{d}{b_d} \rceil, 1, \lceil \frac{n}{b_n} \rceil)$-blocking, and its outermost loop iterates over columns of $A$ to encourage caching of the sparse matrix data and $\hat{A}$.
It does not block the inner dimension since that dimension presents fewer opportunities for cache behavior optimization when working with the CSC sparse matrix format, and it is harder to parallelize over.

The call to \texttt{compute\_kernel} in \Cref{alg:simple} handles multiplying the submatrices produced by outer blocking, and is our next topic of discussion.
The topic of setting block sizes in \Cref{alg:simple} is addressed in  \Cref{section:performance,subsec:parallel_runs}.


\subsection{Compute Kernel (Inner Blocking)}\label{section:ordering}

The main question in the implementation of \texttt{compute\_kernel} is how to handle loop ordering.
For example, \Cref{alg:order} loops over the rows of the left operand ($L$), then over the inner dimension (columns of $L$ and rows of $R$),
In interpreting the pseudocode, one should think of $L$ and $R$ as submatrices of $S$ and $A$ obtained through blocking.
The pseudocode uses dense indexing notation for the sparse block $R$ only to simplify exposition.

\begin{algorithm}[H]
    \caption{Toy compute kernel with $ijk$ loop ordering
    } \label{alg:order}
    \begin{algorithmic}[1]
        \Require{\multiline{$L$\,: dense $d_1 \times m_1$ matrix \\ $ R$\,: sparse $m_1 \times n_1$ matrix \\}}
            \State $G := 0_{\mathbb{R}^{d_1 \times n_1}}$
            \For{$i = 1 \texttt{:} d_1$}
                \For{$j = 1 \texttt{:} m_1$}  
                    \For{$k = 1 \texttt{:} n_1$}  
                        \State $G[i, k] = G[i, k] + L[i, j]R[j, k]$
                    \EndFor
                \EndFor
            \EndFor
            \State \textbf{return} $G$
    \end{algorithmic}
\end{algorithm}

There are six possible choices for this loop ordering.
We list them all below in the special case of 3-by-3 matrices to balance completeness and concreteness.
In the descriptions, $\ell_{ab}$ denotes the $(a,b)^{\text{th}}$ entry of $L$, while $\hat{\ell}_a$ and $\ell_b$ denote the $a^{\text{th}}$ row and $b^{\text{th}}$ column of $L$.
The analogous conventions apply when interpreting $r_{ab}$, $\hat{r}_a$, and $r_b$ for $R$. 


\begin{itemize}
    \item Variants \texttt{ikj} and \texttt{kij}: 
    \begin{align*}
        \begin{pmatrix}
       \hat\ell_1 \\ \hat\ell_2 \\ \hat\ell_3
    \end{pmatrix}\begin{pmatrix}
       r_1 & r_2 & r_3
    \end{pmatrix} &= \begin{pmatrix}
       \hat\ell_1r_1 & \hat\ell_1r_2 & \hat\ell_1r_3 \\
       \hat\ell_2r_1 & \hat\ell_2r_2 & \hat\ell_2r_3 \\
       \hat\ell_3r_1 & \hat\ell_3r_2 & \hat\ell_3r_3
    \end{pmatrix}
    \end{align*}
    Variant \texttt{ikj} corresponds to streaming through $G$ in row major order, 
    and variant \texttt{kij} corresponds to column major.
    \item Variant \texttt{ijk}:
    \begin{align*}
        \begin{pmatrix}
       \ell_{11} & \ell_{12} & \ell_{13} \\
       \ell_{21} & \ell_{22} & \ell_{23} \\
       \ell_{31} & \ell_{32} & \ell_{33}
    \end{pmatrix}\begin{pmatrix}
       \hat{r_1} \\ \hat{r_2} \\ \hat{r_3}
    \end{pmatrix} &= \begin{pmatrix}
       \ell_{11}\hat{r_1} + \ell_{12}\hat{r_2} + \ell_{13}\hat{r_3} \\ 
       \ell_{21}\hat{r_1} + \ell_{22}\hat{r_2} + \ell_{23}\hat{r_3} \\ 
       \ell_{31}\hat{r_1} + \ell_{32}\hat{r_2} + \ell_{33}\hat{r_3} \\
    \end{pmatrix}
    \end{align*}
    \item Variant \texttt{kji}:
    \begin{align*}
        &\begin{pmatrix}
       \ell_1 & \ell_2 & \ell_3
    \end{pmatrix}\begin{pmatrix}
       r_{11} & r_{12} & r_{13} \\
       r_{21} & r_{22} & r_{23} \\
       r_{31} & r_{32} & r_{33}
    \end{pmatrix} \\ 
     = &\begin{pmatrix}
       \sum_{i = 1}^3 r_{i1}\ell_i & 
        \sum_{i = 1}^3 r_{i2}\ell_i & 
        \sum_{i = 1}^3 r_{i3}\ell_i \\
    \end{pmatrix}
    \end{align*}
    \item Variants \texttt{jik} and \texttt{jki}
    \begin{align*}
        \begin{pmatrix}
       \ell_1 & \ell_2 & \ell_3
    \end{pmatrix}\begin{pmatrix}
       \hat{r_1} \\ \hat{r_2} \\ \hat{r_3}
    \end{pmatrix} &= \ell_1\hat{r_1} + \ell_2\hat{r_2} + \ell_3\hat{r_3}
    \end{align*} For each $\ell_i\hat{r_i}$, variant \texttt{jik} updates $G$ in row major order, variant \texttt{jki} updates $G$ in column major order.
\end{itemize}


We can quickly rule out three of these options.
First, note that the first two compute variants (\texttt{ikj} and \texttt{kij}) require noncontiguous random number generation since the dot product $\hat \ell_ir_j$ for some $i, j$ only requires numbers in $\hat \ell_i$ to be generated at locations corresponding to the nonzero entries in $r_j$.
However, because it is difficult to apply vectorization techniques to noncontiguous random number generation, we remove these from consideration.
We can also rule out variant \texttt{ijk}, since summing together rows of $R$ would be inefficient regardless of the sparse matrix format.

In evaluating the remaining loop ordering options we must consider our target architectures.
Some architectures may favor strided accesses and severely penalize random accesses even if the entries are already in fast cache.
These properties can mostly be attributed to prefetching mechanisms and cache prediction algorithms.
Based on this observation, we can divide the algorithms into two cases.

\subsubsection{Architectures that are sensitive to random access}
\begin{figure}[ht]
\begin{center}
\includegraphics[width=0.7\linewidth]{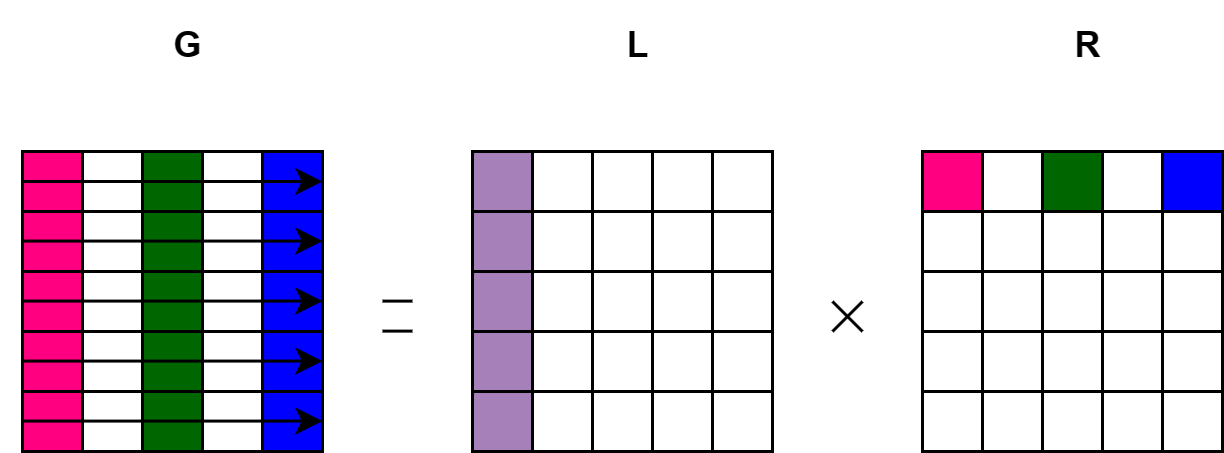} \\
\end{center}
\caption{Variant \texttt{jik}. Updates to $G$ are row-wise.}
 \label{fig:var5}
\vspace{-0.5cm}
\end{figure}

\begin{figure}[ht]
\begin{center}
\includegraphics[width=0.7\linewidth]{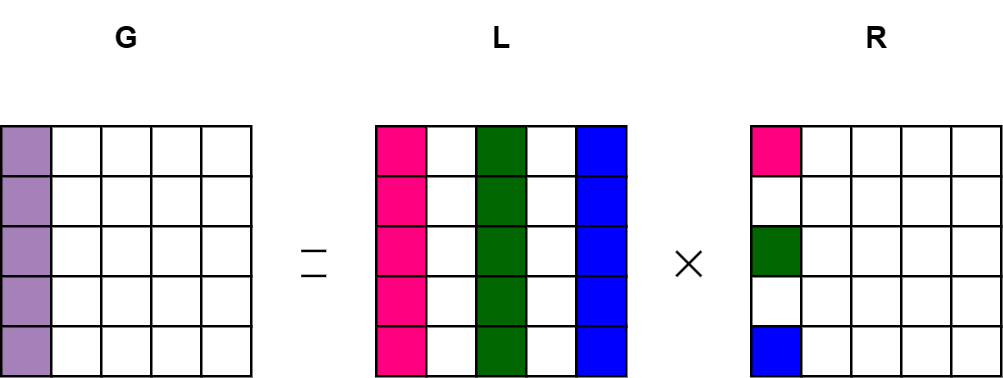} \\
\end{center}
\caption{Variant \texttt{kji} (\Cref{ck:v4}). Each column of $G$ is updated using linear combination of columns of $L$. Sparsity leads to gaps between columns of $L$, but each column is accessed contiguously.
\label{fig:var4}
}
\end{figure}

\begin{figure}[ht]
\begin{center}
\includegraphics[width=0.7\linewidth]{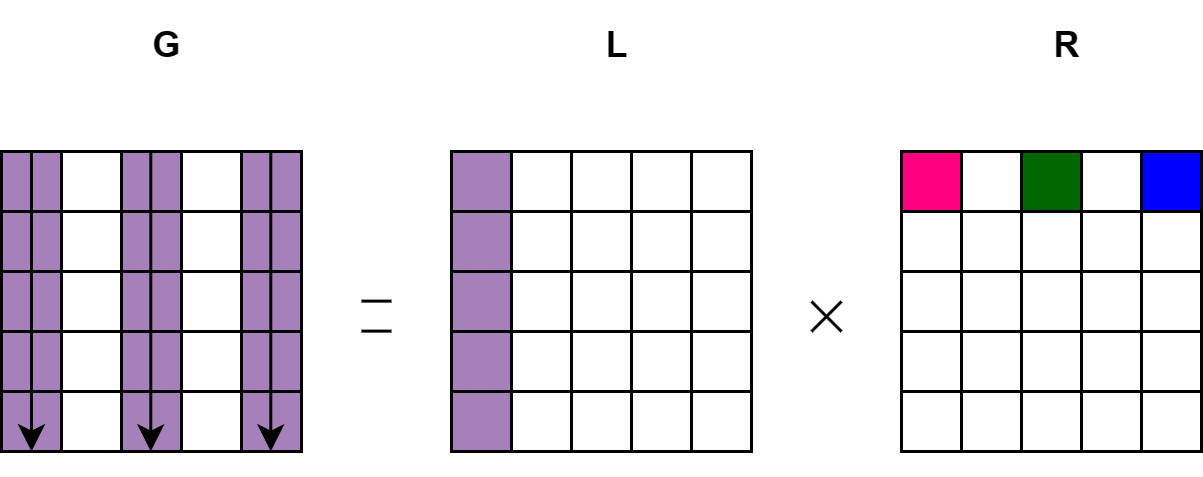} \\
\end{center}
\caption{Variant \texttt{jki} (\Cref{ck:v6}); $G$ is updated column-wise.}
 \label{fig:var6}
\end{figure}

Variant \texttt{jik} (visualized in \Cref{fig:var5}) updates $G$ in row major order. Since the rows of $R$ are sparse these updates would be noncontiguous, and hence we can remove \texttt{jik} from contention.
This leaves us with only \texttt{kji} and \texttt{jki} (visualized in \Cref{fig:var4,fig:var6}).
The difference between the two is that the former requires random access to slices of $L$ while the latter requires random access to slices of $G$.
Note that while the access patterns of both methods depends on the sparsity structure of $R$, they are nevertheless more cache friendly than variant \texttt{jik}.

Variant \texttt{kji} can naturally leverage on-the-fly random number generation.
The pseudocode for this is in \Cref{ck:v4}.
Notice that the algorithm uses dense notation for simplicity.
In practice, we would change the indexing of $A$ according to the appropriate sparse data structure.
Furthermore, we replaced accesses to $S$ with a vector $v$ that we repeatedly overwrite with generated random numbers.
After this modification and assuming that $A$ is given in CSC format, we will have strided access for all three matrices, making variant \texttt{kji} preferable to variant \texttt{jki}. 

This algorithm is relatively oblivious to the sparsity pattern of $A$ as shown in \Cref{tab:extreme_examples}. In addition, it only requires the simple CSC structure for the input. We can also use SIMD instructions to accelerate the computation. 

\begin{algorithm}
    \caption{
    Variant \texttt{kji} with RNG} 
    \label{ck:v4}
    \begin{algorithmic}[1]
        \Require{\multiline{$\hat{A}_{\mathrm{sub}}$\,: dense $d_1 \times n_1$ submatrix of $\hat{A}$, \\
        $A_{\mathrm{sub}}$\,: sparse $m \times n_1$ submatrix of $A$ \\
        \,$r $\,: row offset for $\hat{A}_{\mathrm{sub}}$ as a submatrix of $\hat{A}$. } \vspace{0.1ex} }
            \State{$v = 0_{\mathbb{R}^{d_1}}$}
            \State{$g = $ \texttt{RNG()}} \codecomment{see \Cref{subsec:RNG_choice} }
            \For{$k = 1 \texttt{:} n_1$}
                \For{$j = 1 \texttt{:} m$}  
                    \If{$A_{\mathrm{sub}}[j, k] = 0$}
                        \State{continue}
                    \EndIf
                    \State{$g.\texttt{set\_state(}r,j\texttt{)}$}
                    \State{$g.\texttt{get\_samples(}v\texttt{)}$}\label{numbergen} \Statex{\qquad \quad \codecomment{$v$ is overwritten with $d_1$ random numbers}}
                    \For{$i = 1 \texttt{:} d_1$}  
                         \State{$\hat{A}_{\mathrm{sub}}[i, k] \mathrel{+}= A_{\mathrm{sub}}[j, k] * v[i]$}
                         \Statex{\codecomment{In actual implementations, index according to $A$'s sparse}}
                         \Statex{\codecomment{matrix data structure. CSC is the most suitable here.}}
                    \EndFor
                \EndFor
            \EndFor
            \State \textbf{return} 
    \end{algorithmic}
\end{algorithm}

\subsubsection{Architectures that don't heavily penalize random access}\label{sec:case2}

If the computational efficiency is not drastically impacted by non-strided accesses resulting from the sparsity pattern of $A$ or if random numbers are expensive to generate, then it is worth reconsidering variant \texttt{jki}.
The reason is that this variant can better take advantage of blocking to reduce the number of randomly generated samples.
To see why this might be the case, we can refer back to \Cref{fig:var4,fig:var6}. 
\Cref{fig:var4} will generate an entire column of $L$ for each entry in $R$, whereas \Cref{fig:var6} reuses the same column of $L$ for an entire row of $R$ before regenerating. The pseudocode for variant \texttt{jki} is in \Cref{ck:v6}, which computes $\hat{A}[i\texttt{:}\text{d\_stop},\, j\texttt{:}\text{n\_stop}]$ by summing rank-1 updates.
The downside of \Cref{ck:v6} is that it demands a more sophisticated data structure.
$A$ will need to be first partitioned into vertical blocks, and within each block, the entries will be stored in CSR format.

\begin{algorithm}
    \caption{Variant \texttt{jki} with RNG}\label{ck:v6}
    \begin{algorithmic}[1]
        \Require{\multiline{$\hat{A}_{\mathrm{sub}}$\,: dense $d_1 \times n_1$ submatrix of $\hat{A}$, \\
        $A_{\mathrm{sub}}$\,: sparse $m \times n_1$ submatrix of $A$ \\
        \,$r $\,: row offset for $\hat{A}_{\mathrm{sub}}$ as a submatrix of $\hat{A}$. } \vspace{0.1ex} }
            \State{$v = 0_{\mathbb{R}^{d_1}}$}
            \State{$g = $ \texttt{RNG()}}
            \For{$j = 1 \texttt{:} m$}
                    \If{$A_{\mathrm{sub}}[j, \texttt{:}] = 0_{\mathbb{R}^{n_1}}$}
                        \State{continue}
                    \EndIf
                    \State{$g.\texttt{set\_state}(r, j)$}
                    \State{$g.{\texttt{get\_samples}(v)}$} 
                \For{$k = 1 \texttt{:} n_1$}  
                    \For{$i = 1 \texttt{:} d_1$}  
                         \State{$\hat{A}_{\mathrm{sub}}[i, k] \mathrel{+}= A_{\mathrm{sub}}[j, k] * v[i]$}
                        \Statex{\codecomment{In implementations, index according to the sparse matrix}}
                        \Statex{\codecomment{data structure of $A_{\mathrm{sub}}$ (CSR is the most suitable).}}
                    \EndFor
                \EndFor
            \EndFor
            \State \textbf{return} 
    \end{algorithmic}
\end{algorithm}

\subsection{Parallellism}

Once, the compute kernel variant is chosen, we can pair it with an outer blocking scheme, such as \Cref{alg:simple}, to perform the entire multiplication. We now can consider where parallelism can be inserted to gain the best performance boost.
One approach is to use multithreaded random number generation (line~\ref{numbergen} of \Cref{ck:v4}). This, however, is ineffective since in most cases the multithreading overhead would be greater than the work itself.
A simple and effective approach is to parallelize either of the two loops in \Cref{alg:simple}.
This gives reasonable results, as we show in \Cref{sec:experimental}.

\section{Analysis}

Here we discuss how our methodology can trade off between data movement and  computational cost.
We start with establishing a theoretical bound for the ratio of these quantities under a simplified model that allows for blocking over all three loop indices.
Then we discuss more practical aspects of the computational cost / data movement tradeoff as they pertain to \Cref{alg:simple} (which only blocks the outer loop indices).
Finally, we present an experiment that shows how changing cost of random number generation affects our method when using \Cref{ck:v6} as the compute kernel.
\subsection{Theoretical Peak Performance}
\label{section:performance}


We will use the roofline model \cite{roofline} to roughly estimate the theoretical peak performance of the dense times sparse operation. It's very difficult to account for all the computer architecture factors and practical optimization tricks in our performance model. However, the derivations will provide some grounds for understanding how parameters such as the performance of random number generators can potentially impact the overall performance. 

For modelling purposes, we measure the cost of generating random numbers relative to the cost of memory access. Let $h > 0$ denote the cost of generating one random number, where $h = 1$ means that generation cost is equivalent to a memory access. We consider the case $h<1$, i.e. it is cheaper to compute an entry of $S$ on the fly than to read it from memory, since otherwise it is asymptotically faster to precompute $S$ once, store it in memory, and then 
optimize matrix multiplication $S \cdot A$ in a standard way.

Let $M$ denote the cache size for our analysis. We assume a simple one layer cache model in which matrix entries have to be moved from the main memory into cache before computation. 
In general, it would be difficult to find a formula to describe sparse matrices with arbitrary nonzero patterns. Hence, we do the analysis on a uniformly distributed sparse matrix with a density of $\rho$. In other words, we assume any sub-matrix will also have a density of $\rho$.
For simplicity, we will optimize over blocks of a fixed size.
The submatrix of $S$ will be $d_1 \times m_1$, the submatrix of $A$ will be $m_1 \times n_1$, and (consequently) the submatrix of $\hat{A}$ will be $d_1\times n_1$.
We denote these submatrices by $\hat{A}_{\mathrm{sub}}$, $S_{\mathrm{sub}}$, and  $A_{\mathrm{sub}}$ moving forward.

Our model stipulates that each time we fill the cache we can perform $2\rho d_1m_1n_1$ operations. 
Since $h<1$ and we regenerate entries of $S$ on the fly, $S$ doesn't occupy valuable cache space.
However, we need to account for the cost of generating entries of $S_{\mathrm{sub}}$ in a particular block operation.
Notice that if row $i$ of $A_{\mathrm{sub}}$ has at least one nonzero, then the corresponding column of $S_{\mathrm{sub}}$ (i.e., $d_1$ random numbers) must be generated.
We use a simple model for how often this happens in expectation.
Let $X_i$ be the random variable that equals either 1 or 0, denoting whether row $i$ of $A_{\mathrm{sub}}$ has at least one nonzero, so that $Y = \sum_{i = 1}^{m_1} X_i$ counts the number of rows that have at least one nonzero.
Assuming that each entry of $A$ is nonzero with probability $\rho$, we have 
$\mathbb{E}[Y] = m_1(1 - (1 - \rho)^{n_1})$.
Hence, the expected 
total amount of memory movement plus the total cost of number generation is $2dmn\rho (M + hd_1m_1(1 - (1 - \rho))^{n_1}) / (2\rho d_1m_1n_1)$.

We want to maximize computational intensity (CI).
This is normally defined as the number of computations divided by the total memory movement.
However, in our case, we incorporate the generation of $S$ into data movement cost using the scale factor ``$h$'' described earlier.

We minimize the reciprocal of CI to simplify exposition.
That is, we want to solve the following problem:
\begin{align} 
    \min_{m_1, n_1, d_1}& \quad \frac{dmn (M + hd_1m_1(1 - (1 - \rho)^{n_1}))}{d_1m_1n_1} \notag \\
    \text{subject to}& \quad d_1n_1 + m_1n_1\rho \leq M. \label{optimization:problem}
\end{align}
Our first step is to pretend that $n_1$ is fixed and express our objective function with a new variable $y = d_1m_1$:
\begin{align*}
    &\frac{dmn (M + hd_1m_1(1 - (1 - \rho)^{n_1}))}{d_1m_1n_1} \\
    &= \frac{dmnM}{n_1 y} + \frac{hdmn(1 - (1 - \rho)^{n_1})}{n_1}.
\end{align*}
The optimal choice of $y$ is the largest value that satisfies the cache constraint.
The bound on $y$ is maximized when we set $d_1 = M / (2n_1)$ and $m_1 = M / (2n_1\rho)$. 
Substituting these expressions for $d_1$ and $m_1$ into the objective leaves us with an unconstrained optimization problem:
\begin{align*}
    \min_{n_1} \frac{4n_1\rho dmn}{M} + \frac{dmnh(1 - (1 - \rho)^{n_1})}{n_1}.
\end{align*}
There is no closed form solution for the optimal value of $n_1$.
However, we can make approximations based on the value $\rho$. 

\subsubsection{$\rho$ is small}
\label{subsubsection:smallrho}
If $\rho$ is close to zero, then we have $\frac{1-(1-\rho)^{n_1}}{n_1} \approx \frac{1-(1-n_1\rho)}{n_1} = \rho$. In this case, $n_1 = 1$ would be the optimal value, and give a CI of
\begin{equation}
    2dmn\rho\frac{M}{4dmn\rho + Mdmn\rho h} = \frac{2M}{4 + Mh}.
\end{equation}
In order to achieve peak performance, the CI has to be greater than machine balance, which is defined as the theoretical peak GFLOPS of a machine divided by its bandwidth (bytes of memory transferred per second). Using $B$ to denote machine balance, we estimate the theoretical fraction of 
peak to be $O(\frac{1}{hB})$ if $Mh \gg 4$. Otherwise, if $h$ is adequately small, then the theoretical fraction of peak would be
\begin{align} \label{equation:theoretical_peak}
O\left(\frac{M}{B}\right).
\end{align}
The analysis for the $n_1 = 1$ case also applies to sparse matrices with arbitrary sparsity patterns. For reference, the theoretical fraction of peak for GEMM is $O(\sqrt{M} / B)$, so \Cref{equation:theoretical_peak} is a factor of $\sqrt{M}$ better.

\subsubsection{$\rho$ is big}
On the other hand, if $\rho$ is near 1, then we can assume $1 - (1 - \rho)^{n_1} \approx 1$, 
which would yield $n_1 = \frac{\sqrt{hM}}{2\sqrt{\rho}}$ as the minimizer. 
The theoretical fraction of
peak would then be
\begin{equation}
    \frac{dmn\rho}{2Bdmn\sqrt{\frac{h\rho}{M}}} = \frac{\sqrt{M\rho}}{2B\sqrt{h}}.
\end{equation}

\subsection{Comparing the Algorithms}
\label{sec:compare}
We will now analyze the algorithmic costs, including the cost of introducing new data structures for \Cref{ck:v6}.
Here we use $(b_d, b_n)$ to denote the block sizes, rather than $(d_1, n_1)$.

While \Cref{ck:v4} is able to exploit the reproducibility of random numbers to produce hardware friendly memory accesses, it has no way of reusing the randomly generated numbers.
In other words, it will always generate $d \times \operatorname{nnz}(A)$ random numbers.
This resembles the approach in \Cref{subsubsection:smallrho} and is highly dependent on having a fast RNG.

If we want to sacrifice some benefits of strided 
access in order to reduce the amount of numbers generated, such as the case of \Cref{ck:v6}, then we can cut down the total number of randomly generated entries to $O(\lceil \frac{ndm}{b_n} \rceil)$. This is a worst case estimate. In reality, for a given vertical block of $A$, $A_{\mathrm{sub}} = A[\texttt{:},\,j\texttt{:}  \text{n\_stop}]$, there may exist many rows that only contain zeros.
Suppose that row $r$ of $A_{\mathrm{sub}}$ is zero, then the corresponding column of $S[i \texttt{:} \text{d\_stop}, \texttt{:}]$ need not be generated.
Hence, depending on the sparsity pattern of $A$, one could tune $b_n$ to minimize the number of random variables generated.

\Cref{ck:v6} uses an auxiliary blocked CSR data structure, while \Cref{ck:v4} only requires standard CSC, which we assume is given for free.
Constructing a blocked CSR representation from a CSC representation is equivalent to dividing columns of $A$ into blocks and transposing each block of $A$. The cost of  sequential construction is $O(\lceil \frac{n}{b_n} \rceil m + \nnz(A))$. We can also construct this data structure in parallel by assigning blocks to each thread individually. Given $T$ threads, the parallel construction complexity would be $O(\lceil \frac{n}{T b_n} \rceil m + \max_t\nnz(A_t))$, where $A_t$ represent the blocks assigned to thread $t$. Notice that this construction is memory intensive because each block needs $O(m)$ memory to store the number of elements per row.

%
%
%
%


\subsection{Impact of the distribution of random matrix entries}\label{subsec:impact_of_rng}

Typical low-level RNGs only attempt to sample uniformly at random over some fixed-width integers.
To sample random numbers from some other distribution one has to apply a suitable transformation to these integers.
For instance, to obtain a random number $a \in (-1, 1)$, one can generate a random signed 32-bit integer and divide it by $2^{31}$.
However, the transformation can be much more expensive, particularly for the Gaussian distribution.

\begin{figure}[H]
\begin{center}
\includegraphics[width=1\linewidth,trim=0 5 0 5,clip]{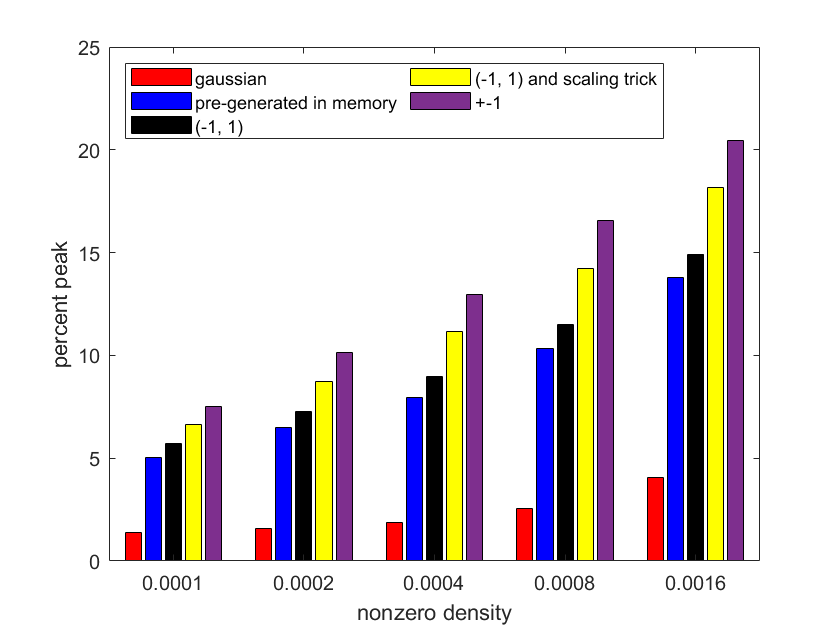}
\vspace{-0.75cm}
\end{center}
\caption{Percent of peak (Perlmutter supercomputer CPU-node) obtained when running \Cref{ck:v6} on sparse uniformly distributed random matrices, as a function of nonzero density. 
Five methods are shown for how we generate entries of $S$: generating Gaussians on the fly, pre-generating $S$ in memory, sampling $(-1, 1)$ on the fly (optionally with a scaling trick), and generating $\pm 1$ on the fly.
All methods except the second generate entries on the fly, and all the numbers are 32 bits except for $\pm 1$, which uses 8-bit integers.
}

 \label{fig:percent_peak}
\end{figure}


\Cref{fig:percent_peak} shows how the speed of \Cref{ck:v6} is affected by using different distributions for the entries of $S$.
It is clear that generating Gaussians on the fly is not practical in this implementation.
For the case of pre-generating $S$ in memory, we don't include generation time.
This is significant because three other methods all achieve a higher percent of peak than pre-generating $S$, which supports our hypothesis that generating suitable random numbers on-the-fly can be used for exceptionally fast sketching.
Finally, we point out the approach in the figure labeled ``$(-1, 1)$ and scaling trick.'' This refers to \textit{effectively} computing $SA$ where the entries of $S$ are sampled iid from $(-1, 1)$ by instead computing $(Sf)(A/f)$ for a scale factor $f = 1 / \texttt{max\_int}$ and representing $Sf$ as the matrix of integers generated from the base RNG.

\section{Implementation Considerations}
\subsection{Programming Language}
We want a programming language that has flexible, easily-to-use SIMD support, and fast random number generation. This motivated us to use the Julia programming language \cite{bezanson2012julia}. 
It has extraordinary support for SIMD operations, including SIMD random number generation and fused multiply-add vector intrinsics with the help of Julia packages \texttt{RandomNumbers.jl} and \texttt{LoopVectorization.jl}. The latter package (written fully in Julia) is capable of generating high performing vectorized kernels that perform on par with MKL \cite{loopvec}. Furthermore, the built-in  
XOR-shift based library is easy to use and has performance that is on par with SIMD RNG libraries written in pure C \cite{c++_xoshiro}. 
To parallelize the outer loops, we simply add the macro \texttt{Threads.@threads} as a prefix to the for-loops.

\subsection{Random Number Generator Choice}\label{subsec:RNG_choice}
In order to produce high performance implementations, we need to carefully select an efficient RNG since we want to be able to generate entries of $S$ at arbitrary locations. We ideally need to be able to recover the state of the RNG in $O(1)$ time. 
\subsubsection{Counter Based RNG}
A counter based RNG (CBRNG), such as one from Random123 library \cite{Random123}, uses independent transformations, so numbers can be generated by simply supplying the generator with a counter, independent of any previous states.
CBRNGs would in theory be the ideal approach since we can simply use the row-column indices of the elements in the matrices as counters.
However, the generators in Random123 were almost a factor five slower than the generators we review next.

\subsubsection{XOR-shift Based RNG}
Xoshiro belongs to a family of XOR-shift based RNGs \cite{xoshiro}.
Given an initial state, Xoshiro produces a sample and modifies the state in the process. The updated state will be used the next time a random number needs to be sampled.

We can attain reproducibility by attaching a unique state to each matrix entry and generating the entry using that state. Xoshiro typically has a tuple of integers as its state, so we can set the state to be the row and column coordinate of the entry. However, the numbers may no longer have the desired statistical properties if we manually change the state for each entry. In addition, the performance can also degrade since a state change interrupts vectorization. Fortunately, the blocked nature of our algorithms implies that we can view each block as an entry. Therefore, we would only need to modify the state once for each block. Essentially, this approach is utilizing blocks as ``checkpoints".
 
 We use a SIMD implementation of Xoshiro.
 Since Xoshiro has sequentially dependent state transformations, block sizes and checkpoint locations will impact reproducibility of the sketch.
 Based on the tests we ran, the quality of the sketches are fine in the context of least squares solver (as measured by effective distortion; see \cite[\S 2]{murray2023randomized}).
 Therefore, unless reproducibility independent of blocking is critical, Xoshiro may be preferable over CBRNGs.

\subsection{The Big Picture}
We have focused on getting the fastest implementation possible.
However, in some contexts, there are constraints that would preclude some of our techniques. For example, RandBLAS \cite{murray2023randomized} 
uses CBRNGs for reproducibility independent of the number of threads used by OpenMP. 
Hence, in order to incorporate this algorithm into the RandBLAS framework, we might opt for a CBRNG.
Future work will be needed to properly compare our algorithms to ideal CBRNG-based algorithms designed for the same sketching task; the outcome of such a comparison could inform changes to the RandBLAS policy on threading-independent random number generation.

\section{Experimental results}\label{sec:experimental}
We test our algorithms on compute nodes of the Frontera (TACC) and Perlmutter (NERSC) supercomputers.
The Frontera node has one Intel Xeon Platinum 8280 (Cascade Lake) CPU at 2.7 GHz and 192 GB of RAM.
The Perlmutter node has two AMD EPYC 7763 (Milan) CPUs at 2.45 GHz and 512 GB of RAM.
All times are given in seconds.

\begin{figure}[H]
\begin{center}
\includegraphics[width=1\linewidth]{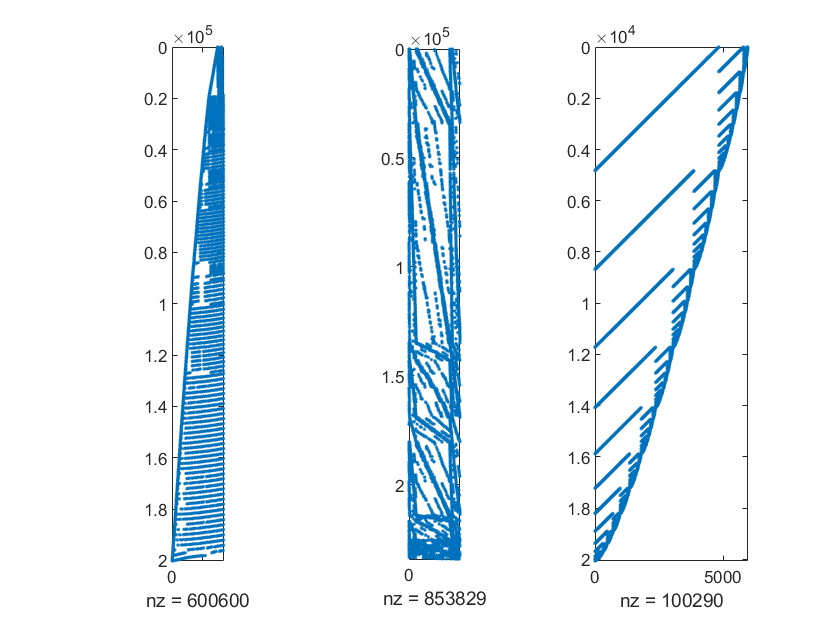}
\end{center}
\caption{Sparsity patterns of some of the matrices used in testing. From left to right, the matrices are shar\_te2-b2, mesh\_deform, cis-n4c6-b4.
 }
 \label{sparsity_pattern}
\end{figure}

Properties of test data for our SpMM experiments are reported in \Cref{matrix:name} and visualized in \Cref{sparsity_pattern}, while test data for the least squares experiments are described in \Cref{matrix:name_extended}. 
In both types of experiments we use a sketch size $d = \gamma n$ for small constants $\gamma$.
In an idealized Gaussian case, this leads the effective distortion of $S$ for $\operatorname{range}(A)$ to converge almost surely to $1/\sqrt{\gamma}$ as $n$ tends to infinity \cite[\S A.1.1]{murray2023randomized}.
To put this in the context of the randomized preconditioning least squares solver in \Cref{sec:exp_least_square}, this results in a preconditioned data matrix with condition number bounded by $(\sqrt{\gamma}+1)/(\sqrt{\gamma}-1)$.
Our SpMM and least squares experiments take $\gamma = 3$ and $\gamma = 2$, respectively.

\begin{table}[H]
\small
\caption{Properties of test data for SpMM benchmarks.
Here, $d = 3n$ is the number of rows in $S$, and $(m, n)$ are the dimensions of $A$.} \label{matrix:name}
\begin{center}
\resizebox{\columnwidth}{!}{%
\begin{tabular}{cccccccc}
\toprule[1.2pt]
Matrices               &  $d$      &  $m$ & $n$ & nnz($A$) & density  \\ \midrule[1.2pt]
mk-12  &   4455      & 13860 & 1485     & 41580 & 2.02E-03 \\ 
ch7-9-b3    &   52920      & 105840 & 17640      & 423360 & 2.27E-04 \\ 
shar\_te2-b2   &  51480     & 200200 & 17160     & 600600 &  1.75E-04 \\ 
mesh\_deform  &   28179     & 234023 & 9393   & 853829 & 3.88E-04  \\ 
cis-n4c6-b4 &  17910    & 20058 & 5970   & 100290 & 8.38E-04 \\ \bottomrule[1.2pt]               
\end{tabular}}
\end{center}
\end{table}

\subsection{Sequential Runs}

\Cref{result:csrrow:tacc} compares our implementation of \Cref{ck:v4} to several library implementations of SpMM on Frontera.
MKL timings use CSR for $A$ and row major storage for $S$ since MKL only supports sparse-times-dense. (Hence, the operation and storage are transposed.) 
Times for \Cref{ck:v4} are reported in two cases: when sampling uniformly from $(-1, 1)$ and when sampling uniformly from $\{+1, -1\}$.
The latter case is cheaper, but may decrease the quality of the sketch.
In both cases, the data shows that our algorithm significantly outperforms the standard libraries.

\begin{table}[ht]
\small
\caption{Timing comparison of \Cref{ck:v4} against several library implementations on Frontera. $b_n = 500, b_d = 3000$.} 
\label{result:csrrow:tacc}
\begin{center}
 \begin{tabular}{cccccc}
\toprule[1.2pt]
\multirow{2}{*}{Matrices} & \multirow{2}{*}{MKL} & \multirow{2}{*}{Eigen} & \multirow{2}{*}{Julia} & \multicolumn{2}{c}{\Cref{ck:v4}} \\ 
 & & & & $(-1, 1)$ & $(\pm 1)$ \\ \midrule[1.2pt]
mk-12  & 0.137 & 0.145   & 0.118    & 0.070 & 0.0501   \\ 
ch7-9-b3  & 16.43 & 16.58   & 14.86    & 7.74  & 5.89     \\ 
shar\_te2-b2  & 21.93 & 22.05   &  27.59   & 10.20  & 7.63   \\ 
mesh\_deform  & 15.82 & 16.08  & 14.99   & 8.65  &  5.74   \\ 
cis-n4c6-b4 & 1.351 & 1.36  & 1.18    & 0.74  & 0.531   \\ \bottomrule[1.2pt]               
\end{tabular}
\end{center}
\end{table}

\begin{table}
\small
\caption{Sample time (time to generate the random numbers $(-1, 1)$) and total SpMM time (sample time plus multiplication) on Frontera. $b_n = 500, b_d = 3000$.}\label{tab:breakdown-simple-tacc}
\begin{center}
\begin{tabular}{cccc}
\toprule[1.2pt]
Matrices & Algorithm & total time & sample time \\ \midrule[1.2pt]
mk-12  & \multirow{5}{*}{\Cref{ck:v4}} &  0.076 & 0.036   \\ 
ch7-9-b3 & & 8.34 & 4.07  \\ 
shar\_te2-b2 & & 11.03 & 5.63 \\ 
mesh\_deform & & 9.26 & 4.40  \\ 
cis-n4c6-b4 & & 0.786 & 0.325 \\ \midrule[1.2pt]     
mk-12 & \multirow{5}{*}{\Cref{ck:v6}} & 0.085 & 0.02  \\ 
ch7-9-b3 & & 11.06 & 2.42 \\
shar\_te2-b2 & & 14.43 & 3.84 \\ 
mesh\_deform & & 8.14 & 2.47 \\ 
cis-n4c6-b4 & & 0.924 & 0.157 \\ 
\bottomrule[1.2pt]               
\end{tabular}
\end{center}
\end{table}


\Cref{tab:breakdown-simple-tacc} deals with \Cref{ck:v4,ck:v6}. It reports both the time needed to generate the random numbers and the total runtime time.
We note that the total times are 
slightly higher than those reported in \Cref{result:csrrow:tacc} since the timer creates additional overhead.
This comparison shows that \Cref{ck:v4}, in general, performs better than \Cref{ck:v6} on Frontera.
Hence, this is the case where architecture is sensitive to random access.

\begin{table}
\label{}
\small
\caption{Benchmark results on Perlmutter. Time spent on format conversion for \Cref{ck:v6} is listed separately from its compute time.
$b_n = 1200, b_d = 3000$.} 
\label{result:csrrow:perlmutter}
\begin{center}
\resizebox{\columnwidth}{!}{%
\begin{tabular}{cccccc}
\toprule[1.2pt]
\multirow{2}{*}{Matrices} & \multirow{2}{*}{Julia}  & \multirow{2}{*}{Eigen}  & \multicolumn{2}{c}{\Cref{ck:v6}}  & \multirow{2}{*}{format conversion}  \\ 
& & & $(-1, 1)$ & $(\pm 1)$ &  \\ \midrule[1.2pt]
mk-12  & 0.054    & 0.0662 &  0.0498 & 0.0431  & 0.0026   \\ 
ch7-9-b3 & 6.44    & 7.72 & 6.32  & 5.40 & 0.059   \\ 
shar\_te2-b2 & 10.13   & 11.75  & 8.60  & 7.10  & 0.095   \\
mesh\_deform & 6.24   & 7.40 & 5.47  &  4.47  & 0.098  \\
cis-n4c6-b4 & 0.519    & 0.623 & 0.513  & 0.453  & 0.005   \\ \bottomrule[1.2pt]               
\end{tabular}}
\end{center}
\vspace{-0.5cm}
\end{table}

\Cref{result:csrrow:perlmutter} shows our benchmarking results on Perlmutter, demonstrating the superiority of our implementations on a different CPU architecture (AMD).
Since \Cref{ck:v6} requires a special sparse storage format, we also report the time it takes to convert from CSC to that format.
In most cases, this conversion cost is cheap compared to that of the computation. 
Times for the AMD AOCL sparse package are not included because it was inefficient for the storage format of interest.

\begin{table}

\caption{Sample time (time to generate the random numbers $(-1, 1)$) and total SpMM time (sample time plus multiplication) on Perlmutter. $b_n = 1200, b_d = 3000$.}
\label{tab:breakdown-simple-perlmutter}
    \begin{center}
    \begin{tabular}{cccc}
\toprule[1.2pt]
Matrices &  Algorithm & total time & sample time \\ \midrule[1.2pt]
mk-12  & \multirow{5}{*}{\Cref{ck:v4}}  & 0.0627 & 0.034 \\ 
ch7-9-b3 & & 7.37 & 3.90 \\ 
shar\_te2-b2 & & 9.89 & 5.40\\ 
mesh\_deform & & 7.68 & 4.21\\
cis-n4c6-b4 & & 0.628 & 0.312\\ \midrule[1.2pt]   
mk-12  & \multirow{5}{*}{\Cref{ck:v6}} & 0.0520 & 0.0142\\ 
ch7-9-b3 & & 6.60 & 2.09 \\
shar\_te2-b2 & & 9.04 & 3.64\\
mesh\_deform & & 5.73 & 2.35\\
cis-n4c6-b4 & & 0.532 & 0.120\\\bottomrule[1.2pt]   
\end{tabular}
\end{center}
\end{table}



\Cref{tab:breakdown-simple-perlmutter} gives the runtime breakdown for \Cref{ck:v4,ck:v6} on Perlmutter with random matrix entries drawn from $(-1, 1)$.
The times in \Cref{tab:breakdown-simple-perlmutter} are slower due to overhead from the timer. 
Here, \Cref{ck:v6} performs better than \Cref{ck:v4}, which is the opposite of what happened on Frontera.
Hence, Perlmutter is an example case where the time saved during the process of generating the random numbers can cover up the penalty incurred by random access. 

We also used \texttt{STREAMBenchmark.jl} to benchmark the memory bandwidth of both Frontera and Perlmutter. The benchmark essentially tests the rate at which the machine can copy a vector. Even though in the actual sketching algorithm the memory access pattern will not be as regular as accessing a strided vector, this benchmark still provides a rough idea of the capabilities of the machine. 
In general, Perlmutter has better bandwidth, but Frontera is faster at generating short random vectors (i.e. length of 10000).
Since our sketching algorithm uses blocking, it will only require generation of short vectors. Therefore, generating numbers in the context of blocking is more efficient on Frontera than on Perlmutter. As mentioned in \Cref{sec:compare}, the performance of \Cref{ck:v4} is more dependent on the speed of RNG than \Cref{ck:v6}. This could explain 
why \Cref{ck:v4} is faster than \Cref{ck:v6} on Frontera, whereas the opposite holds for Perlmutter. In contrast, Perlmutter's cache behavior, prefetching mechanism, and data movement rate is likely superior, so it falls under the case described in \Cref{sec:case2}.

\begin{table}
\caption{Comparison of algorithms on synthetic examples with exotic sparsity patterns and comparable nonzero density. Conversion time not included in compute time.}\label{tab:extreme_examples}
\begin{center}
\begin{tabular}{ccccc}
\toprule[1.2pt]
Problem   & Algorithm & conversion time  & compute time \\ \midrule[1.2pt]
\multirow{2}{*}{Abnormal\_A} & \Cref{ck:v4} & N/A & 8.56\\
 & \Cref{ck:v6} & 0.035 & 4.40 \\ \midrule[1.2pt]          
\multirow{2}{*}{Abnormal\_B}  & \Cref{ck:v4} & N/A & 8.51\\
 & \Cref{ck:v6} & 0.085 & 6.10 \\ \midrule[1.2pt]        
 \multirow{2}{*}{Abnormal\_C}  & \Cref{ck:v4} & N/A & 8.46\\
 & \Cref{ck:v6} & 0.056 & 9.43 \\ \bottomrule[1.2pt] 
\end{tabular}
\end{center}
\end{table}


\Cref{tab:extreme_examples} reports results for both algorithms on a few synthetic examples with exotic sparsity patterns.
\texttt{Abnormal\_A} is constructed so that every $1000^{\text{th}}$ row is dense and all other rows are zero. \texttt{Abnormal\_B} has approximately $\frac{2998}{3000}$ of the total nonzeros concentrated around the middle third vertical block. \texttt{Abnormal\_C} is set up so that every $1000^{\text{th}}$ column is dense and all other columns are zero.
All matrices have dimensions $m = 100000, n = 10000$ and a nonzero density of approximately $10^{-3}$.
In all cases, the entries of $S$ are iid uniform over $(-1, 1)$. 

There are two key observations from \Cref{tab:extreme_examples}.
First, \Cref{ck:v6} is able to achieve much better performance than \Cref{ck:v4} for some sparsity patterns.
Second, \Cref{ck:v4} exhibits consistent performance even for very different patterns.
Indeed, time required by \Cref{ck:v6} more than doubled going from \texttt{Abnormal\_A} to \texttt{Abnormal\_C}.
The reason for the performance gap stems from how \Cref{ck:v6} is based on taking outer products (i.e. multiplying one column of the sketching matrix with one row of the sparse matrix). This makes it perform poorly when the sparse matrix has most of its elements stored contiguously in columns.
Hence, when dealing with matrices that feature a wide variety of sparsity patterns, \Cref{ck:v4} may be preferable to \Cref{ck:v6}.

As a final note on serial algorithm runs, we note that one can get upper bounds on performance by replacing each randomly generated entry of $S$ with ``junk'' (e.g., a number computed from simple addition).
In informal experiments this provided for a factor 2x speed up on matrices such as \texttt{shar\_te2-b2}.
This suggests that a fast RNG implemented in hardware would be impactful in dense sketching of sparse matrices.

\begin{table*}
\caption{Parallel scaling results of \Cref{ck:v4,ck:v6} on \texttt{shar\_te2-b2} with $d = 3n$ (Perlmutter). {\bf setup1} uses $b_n=1200, b_d=3000$ whereas {\bf setup2} uses $b_n=100, b_d=15000$. Time for \Cref{ck:v6} doesn't include format conversion.}
\label{tab:parallel}
\vspace{-0.25cm}
\begin{center}
\resizebox{0.6 \textwidth}{!}{%
\begin{tabular}{llllllllll}
    \toprule[1.2pt]
        threads & \multicolumn{2}{c}{\Cref{ck:v6}}  & \multicolumn{2}{c}{\Cref{ck:v4}} & \multicolumn{2}{c}{\Cref{ck:v6}} & \multicolumn{2}{c}{\Cref{ck:v4}} \\ 
        ~ & \multicolumn{2}{c}{setup1} & \multicolumn{2}{c}{setup1} & \multicolumn{2}{c}{setup2} & \multicolumn{2}{c}{setup2} \\ 
        ~ & time(s) & GFlops & time(s) & GFlops & time(s) & GFlops & time(s) & GFlops  & ~ \\ \midrule[1.2pt]
        1 & 8.66 & 7.14 & 9.00 & 6.87 & 8.42 & 7.35 & 8.88 & 6.96  & ~ \\ 
        2 & 5.06 & 12.23 & 5.16 & 11.98 & 4.88 & 12.68 & 4.52 & 13.68  & ~ \\ 
        4 & 2.72 & 22.70 & 2.63 & 23.47 & 2.51 & 24.59 & 2.50 & 24.75  & ~ \\ 
        8 & 2.07 & 29.89 & 1.98 & 31.22 & 1.55 & 39.88 & 1.35 & 45.80  & ~ \\ 
        16 & 2.34 & 26.42 & 1.14 & 54.08 & 1.37 & 45.05 & 0.83 & 74.76  & ~ \\ 
        32 & 2.01 & 30.74 & 0.92 & 67.33 & 0.80 & 77.22 & 0.62 & 100.29 & ~ \\ \bottomrule[1.2pt]
    \end{tabular}}
\end{center}
\end{table*}


\subsection{Parallel Runs}\label{subsec:parallel_runs}

One of the main obstacles to parallel sparse algorithms is the increasing cost of memory traffic that scales up with the number of threads.
Eventually when the memory bandwidth is saturated, the parallel algorithm becomes memory-bound and performance will degrade.
In our context, this challenge can be approached by tuning blocking parameters $(b_n, b_d)$.

Heuristically, setting $b_d$ to larger values and decreasing $b_n$ works to increase the computation to memory movement ratio for the input data matrix $A$.
This highly rectangular blocking structure offloads more ``data access'' cost to the dense sampling matrix $S$.
However, since the entries of $S$ are generated on the fly to a small reserved location, the data access cost is replaced by operations that are less memory intensive.
This decreases the rate at which the memory gets saturated and leads to better scalability.

To make this concrete, \Cref{tab:parallel} shows parallel scalability of \Cref{ck:v4,ck:v6} with two blocking configurations.
The results show \Cref{ck:v4} is preferable in parallel settings due to its potential for superior scaling.
In addition, the format conversion needed by \Cref{ck:v6} is more expensive in the more scalable of the two blocking schemes considered.

\subsection{Applying Sketching to Solve Least Squares Problems}
\label{sec:exp_least_square}

Now we turn to experiments on Perlmutter with a randomized least squares solver based on our sketching method.
The matrices used in these experiments are described in \Cref{matrix:name_extended}.
We removed 158 empty columns from ``specular" and 54 empty rows from ``connectus". 
Test matrices that have $n \gg m$ are transposed.\footnote{In practice, these matrices could arise directly in underdetermined least squares problems. Underdetermined problems can be handled with minor modifications relative to the overdetermined problems we consider.}
We set $b$ in \eqref{eq:least_square} to a random vector in the range of $A$ plus a random Gaussian vector drawn from $\mathcal{N}(0, I)$.

\begin{table*}
\caption{Properties of matrices used in least squares experiments. size($A$) gives the dimensions of $A$ before any transposition, cond($A$) denotes the condition number of $A$, and mem($A$) gives the Mbytes needed to store $A$ in CSC format.} 
\label{matrix:name_extended}
\vspace{-0.25cm}
\begin{center}
\resizebox{0.6 \textwidth}{!}{%
      \begin{tabular}{cccccccc}
\toprule[1.2pt] 
$A$ &  \multicolumn{2}{c}{size($A$)} & nnz($A$) & cond($A$) & cond($AD$) & mem($A$) & density  \\ \midrule[1.2pt] 
rail2586 & 2586 & 923269  &  8011362 &  496.00 &  263.44   & 135.57 & 3.36E-03 \\ 
spal\_004 & 10203 &  321696 & 46168124  & 39389.87 &  1147.79    & 741.26 & 1.41E-02\\
rail4284  &   4284 &  1096894 & 11284032  & 399.78 &  333.87   &  189.32 & 2.40E-03\\ 
rail582  & 582 &  56097 & 402290   & 185.91 &   180.49     & 6.89 & 1.23E-02\\   
specular & 477976 & 1442 & 7647040 & 2.31E+14 &     29.85      & 122.37 & 1.00E-02\\ 
connectus & 458 &  394792 &  1127525 & 1.27E+16 &       1.28E+16     & 21.20 & 5.58E-03\\
landmark  &  71952 &  2704 & 1146848 &  1.39E+18 &    2.30E+17      & 18.37 & 5.89E-03\\ \bottomrule[1.2pt] 
\end{tabular}}
\end{center}
\vspace{-0.25cm}
\end{table*}

\subsubsection{Algorithms}
Our randomized least squares solver follows the \textit{sketch and precondition} (SAP) paradigm \cite{LSRN, murray2023randomized}.
This approach performs a matrix decomposition (QR or SVD) on a sketch $\hat{A} = SA$ to get a preconditioner, and then uses a classical iterative solver.
We use LSQR \cite{LSQR} as the iterative solver and compute $SA$ by running \Cref{ck:v4} with 32 threads.
We note that SAP based on SVD of $SA$ is intended when the original problem has singular values that are near zero.
Naturally, the SVD is much more expensive than QR, and alternative decompositions may be preferable in some contexts.
Our SVD-based SAP implementation drops singular values that are smaller than $\sigma_{\max}(SA) / 10^{12}$.

We compare SAP to two classical algorithms.
The first is LSQR with a simple diagonal preconditioner (henceforth, ``LSQR-D''), constructed so that $D_{ii} = 1/\|A_i\|_2$, where $A_i$ is the $i^{\text{th}}$ column of the input matrix $A$ (after transposing if necessary).
If $\|A_i\|_2 \leq \epsilon\sqrt{n}\max_i \|A_i\|_2$, then we set $D_{ii} = 1$.
The second classical algorithm is a direct method based on SuiteSparseQR (accessed with Julia's backslash solver $A \backslash b$).

\subsubsection{Performance Data}
\begin{figure}
\begin{center}
\includegraphics[width=1\linewidth]{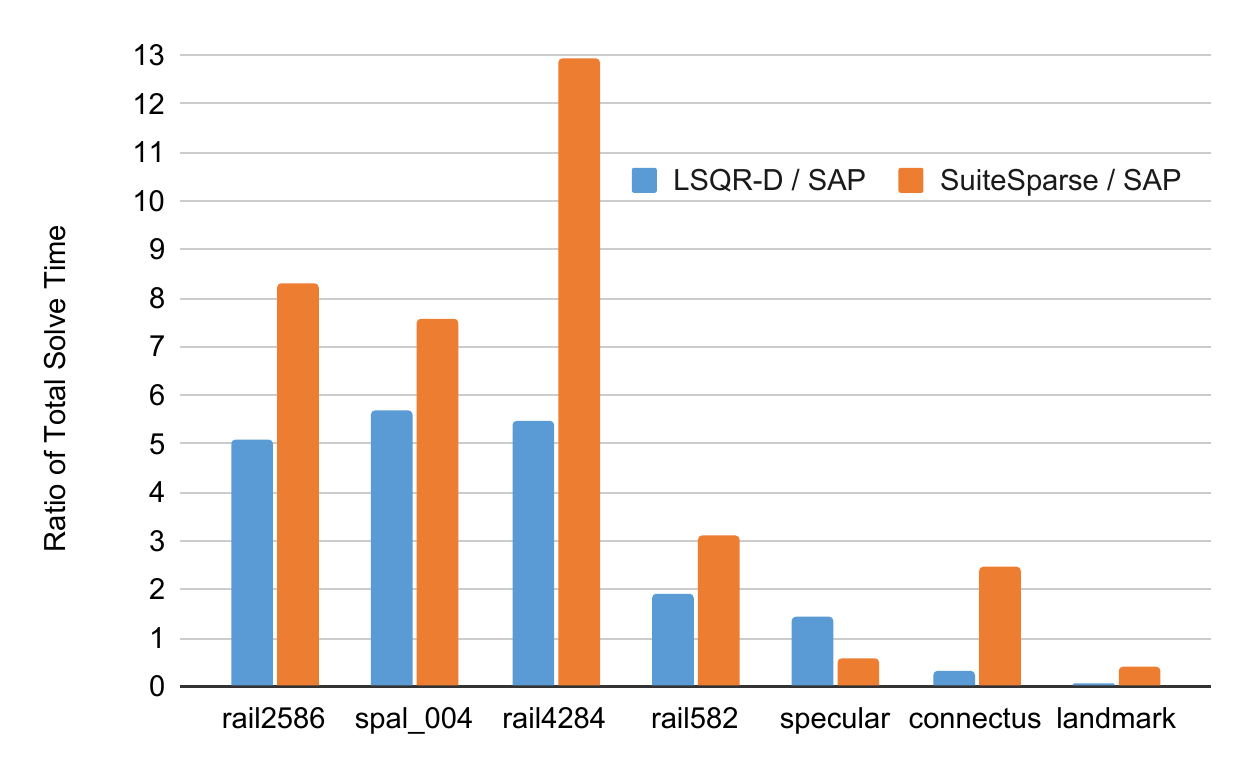}\\
\end{center}
\caption{Comparison of SAP to LSQR-D and SuiteSparse. Let $t_1, t_2, t_3$ be the time it takes to solve least squares problems using LSQR-D, SAP and SuiteSparse, respectively. The figure plots $t_1 / t_2$ (blue) and $t_3 / t_2$ (orange).}
 \label{fig:ratio}
\end{figure}

\Cref{table:solve_result} gives runtime data for the sparse least squares solvers described above.
Qualitative messages from this table can be gleaned from \Cref{fig:ratio}.
In particular, the data shows speedups of up to 13x over SuiteSparse and 5x over LSQR-D.
While SAP is not always the fastest, there is only one matrix (``landmark'') for which it is slower than both LSQR-D and SuiteSparse.
Moreover, the number of iterations required by SAP's iterative solver exhibits practically no variation in comparison to LSQR-D.
Therefore the runtime of SAP is very predictable for different matrix structures, and depends primarily on the cost of computing and factoring $SA$.

It is well known that direct methods can produce more accurate solutions than iterative methods.
Therefore for fair comparison we ran LSQR until its internal (preconditioned) error metric fell below $10^{-14}$.
LSQR employs two such metrics, both motivated by how they bound backward error of an approximate solution.
In our context the relevant error metric for a candidate solution $x$ is
\[
\text{Error}(x) = \frac{\|A^{\top}(Ax - b)\|_2 }{\|A\|_{\text{F}}\|Ax - b\|_2}.
\]
\Cref{table:lsquality} lists this error metric for the solutions returned by each of the three algorithms.
Remarkably, the accuracy of results from SAP exhibit even less variation than the accuracy of results from SuiteSparse.

\begin{table}[H]
\caption{Runtime and iteration count data for sparse least squares solvers. The  ``time (s)'' column for SAP refers to total runtime, not just time for factoring the sketch and running LSQR.
}
\label{table:solve_result}
\begin{center}
\resizebox{\columnwidth}{!}{%
    \begin{tabular}{ccccccc}
    \toprule[1.2pt]
     & \multicolumn{2}{c}{LSQR-D} &  \multicolumn{3}{c}{SAP-QR} &  SuiteSparse  \\ 
$A$ & time (s) & iter & sketch (s) & time (s) & iter & time (s) \\
\midrule[1.2pt] 
      rail2586 & 24.23 & 1412 & 1.17 & 4.78 & 87 & 39.75  \\ 
        spal\_004 & 381.23 & 4830 & 11.48 & 66.99 & 80 & 508.41  \\ 
        rail4284 & 63.00 & 2562 & 2.65 & 11.52 & 88 & 149.27  \\ 
        rail582 & 0.34 & 477 & 0.07 & 0.18 & 80 & 0.55 \\ \bottomrule[1.2pt]
    \end{tabular}}
\end{center}

\begin{center}
\resizebox{\columnwidth}{!}{%
\begin{tabular}{ccccccc}
\toprule[1.2pt] 
 & \multicolumn{2}{c}{LSQR-D} &  \multicolumn{3}{c}{SAP-SVD} &  SuiteSparse  \\ 
$A$ & time (s) & iter & sketch (s) & time (s) & iter & time (s) \\
\midrule[1.2pt] 
 specular & 4.92 & 351 & 0.35 & 3.43 & 79 & 2.04  \\ 
        connectus & 0.19 & 73 & 0.13 & 0.60 & 77 & 1.46  \\
        landmark & 0.80 & 462 & 0.11 & 9.61 & 80 & 3.74 \\  \bottomrule[1.2pt] 
\end{tabular}}
\end{center}
\end{table}

\begin{table}[H]
\caption{Numerical error in computed least squares solutions. Metrics for LSQR-D and SAP deviate slightly from LSQR's stopping tolerance of 1E-14, since LSQR measures error with respect to the precondtioned systems.}
\label{table:lsquality}
    \begin{center}
\begin{tabular}{cccc}
\toprule[1.2pt] 
$A$ & LSQR-D & SAP  & SuiteSparse \\ \midrule[1.2pt] 
  rail2586 & 2.17E-14 & 3.24E-15 & 1.82E-15  \\ 
        spal\_004 & 3.36E-14 & 1.29E-15 & 1.03E-16  \\
        rail4284 & 1.59E-14 & 2.55E-15 & 1.73E-15  \\ 
        rail582 & 1.28E-14 & 5.21E-15 & 7.02E-16  \\ 
        specular & 7.16E-15 & 3.30E-15 & 1.62E-14  \\ 
        connectus & 2.80E-15 & 5.33E-15 & 4.48E-15  \\ 
        landmark & 5.65E-15 & 2.64E-15 & 5.30E-16 \\  \bottomrule[1.2pt] 
\end{tabular}
\end{center}
\end{table}

Having made comparisons in terms of runtime and accuracy, \Cref{table:memory} speaks to \textit{memory efficiency}.
It reports the amount of memory (in Mbytes) required to store $A$ in its CSC representation and the extra memory required by SAP and SuiteSparse (LSQR-D requires essentially no extra memory).
The memory required by SAP is easy to predict, since it only needs a $2n$ by $n$ matrix to store the sketch $SA$.
We measure the memory requirement of SuiteSparseQR by looking at the memory usage of the resulting factors. 
The data shows that SAP required between 7x and 130x less memory than SuiteSparse on these examples.

\begin{table}
    \caption{Memory requirements for two algorithms and for storing the least squares data matrix. All units in Mbytes.}
\label{table:memory}
    \begin{center}
\begin{tabular}{cccc}
\toprule[1.2pt] 
$A$ &  SAP &  SuiteSparse & mem($A$) \\ \midrule[1.2pt] 
rail2586 & 107.00 & 15950.11 & 135.57  \\ 
        spal\_004 & 1665.62 & 49807.51 & 741.26  \\ 
        rail4284 & 293.64 & 38959.24 & 189.32  \\ 
        rail582 & 5.42 & 218.94 & 6.89  \\ 
        specular & 33.27 & 984.10 & 122.37  \\
        connectus & 3.36 & 769.55 & 21.2  \\ 
        landmark & 116.99 & 850.54 & 18.37 \\  \bottomrule[1.2pt] 
\end{tabular}
\end{center}
\end{table}

%
%
%
    

\section{Conclusion and Future Work}
In this paper, we introduce fast SpMM algorithms for the case where the dense matrix is random.
We also provide a theoretical analysis for our algorithms and incorporate them into a least squares solver.
One potential future direction is to extend our theoretical analysis to sparse matrices with non-uniform sparsity patterns.
While a general analysis for arbitrary sparsity pattern may not exist, there are certainly other well-behaved patterns that can be analyzed.
The other direction would be to develop specialized hardware for RNG to further improve the speed of sketching sparse matrices.

\section*{Acknowledgments}

T.Y.\ was supported by the National Science Foundation Graduate Research Fellowship Program under Grant No.\ 2146752.
R.M.\ and J.D.\ were partially funded by an NSF Collaborative Research Framework:
Basic ALgebra LIbraries for Sustainable Technology with Interdisciplinary
Collaboration (BALLISTIC), including the University of California at
Berkeley (NSF Grant No. 2004763) and the International Computer Science Institute (NSF Grant No.\ 2004235).
Any opinions, findings, and conclusions or recommendations expressed in this material are those of the author(s) and do not necessarily reflect the views of the National Science Foundation.

This research is funded in part by the Advanced Scientific Computing Research (ASCR) program within the
Office of Science of the DOE under contract number
DE-AC02-05CH11231. 
We used resources of the NERSC supported by the Office of Science of the DOE under Contract No. DE-AC02-
05CH11231.
R.M.\ was partially supported by Laboratory Directed Research and Development (LDRD) funding from Berkeley Lab, provided by the Director, Office of Science, of the U.S.\ Department of Energy under Contract No.\ DE-AC02-05CH11231.

Sandia National Laboratories is a multimission laboratory managed and operated by National Technology \& Engineering Solutions of Sandia, LLC, a wholly owned subsidiary of Honeywell International Inc., for the U.S. Department of Energy’s National Nuclear Security Administration under contract DE-NA0003525.
This paper describes objective technical results and analysis. Any subjective views or opinions that might be expressed in the paper do not necessarily represent the views of the U.S. Department of Energy or the United States Government.

\bibliographystyle{IEEEtran}
\bibliography{bibtex/bib/biblio.bib}

\end{document}